\documentclass[conference]{IEEEtran}

\usepackage{cite}
\usepackage{textgreek}

\ifCLASSINFOpdf
 \usepackage[pdftex]{graphicx}
 \graphicspath{{../images/}}
 \DeclareGraphicsExtensions{.pdf,.jpeg,.png}
\else
 \usepackage[dvips]{graphicx}
 \graphicspath{{../images/}}
 \DeclareGraphicsExtensions{.eps}
\fi

\usepackage{amsmath}
\usepackage{algorithmic}

\usepackage{array}

\ifCLASSOPTIONcompsoc
 \usepackage[caption=false,font=normalsize,labelfont=sf,textfont=sf]{subfig}
\else
 \usepackage[caption=false,font=footnotesize]{subfig}
\fi

\usepackage{fixltx2e}
\usepackage{stfloats}

\usepackage{url}

\newcommand\DOECopyrightFootnote[1]
{
 \begingroup
 \renewcommand\thefootnote{}\footnote{#1}%
 \addtocounter{footnote}{-1}%
 \endgroup
}

\begin{document}

\title{Havens: Explicit Reliable Memory Regions for HPC Applications}

\author{
\IEEEauthorblockN{Saurabh Hukerikar, Christian Engelmann}
\IEEEauthorblockA{Computer Science and Mathematics Division\\
Oak Ridge National Laboratory\\ 
Oak Ridge, TN, USA\\
Email: \{hukerikarsr, engelmannc\}@ornl.gov}}


\maketitle

\begin{abstract}
Supporting error resilience in future exascale-class supercomputing systems is a critical challenge. Due to transistor scaling trends and increasing memory density, scientific simulations are expected to experience more interruptions caused by transient errors in the system memory. Existing hardware-based detection and recovery techniques will be inadequate to manage the presence of high memory fault rates.  

In this paper we propose a partial memory protection scheme based on region-based memory management. We define the concept of regions called \textit{havens} that provide fault protection for program objects. We provide reliability for the regions through a software-based parity protection mechanism. Our approach enables critical program objects to be placed in these havens. The fault coverage provided by our approach is application agnostic, unlike algorithm-based fault tolerance techniques. 
\end{abstract}



\DOECopyrightFootnote{This work was sponsored by the U.S. Department of Energy's Office of Advanced Scientific Computing Research. This manuscript has been authored by UT-Battelle, LLC under Contract No. DE-AC05-00OR22725 with the U.S. Department of Energy. The United States Government retains and the publisher, by accepting the article for publication, acknowledges that the United States Government retains a non-exclusive, paid-up, irrevocable, world-wide license to publish or reproduce the published form of this manuscript, or allow others to do so, for United States Government purposes. The Department of Energy will provide public access to these results of federally sponsored research in accordance with the DOE Public Access Plan (http://energy.gov/downloads/doe-public-access-plan).}

%
\IEEEpeerreviewmaketitle

\section{Introduction}

The management of faults and errors is a critical concern for the next generation of high-performance computing systems. With transistor scaling, the computing devices fabricated using miniaturized device feature sizes face serious reliability threats. The presence of frequent faults in the hardware, and the resulting errors, will have catastrophic effects on the scientific applications. The applications may not be able to run to completion, or may complete with wrong results due to undetected errors. Therefore, effective fault resilience techniques will be critical for the productive and efficient use of future exascale-class HPC systems \cite{DARPA:ExascaleResilienceStudy}. 

Soft errors in memory are expected to increase dramatically due to increased memory densities in modern HPC systems. Traditionally, reliable memory systems have been equipped with hardware-based ECC, which is capable of correcting single bit errors and detecting double bit errors (SECDED) \cite{Bossen:1980}. However scaling trends, which shrink transistor sizes and reduce supply voltage, result in reduction of charge stored per device, and a correspondingly higher exposure to soft error events. Furthermore, exascale systems are expected to embrace more complex and denser memory hierarchies, and utilize more diverse memory technologies in pursuit of higher memory capacity, and better energy efficiency and performance \cite{DARPA:ExascaleTechStudyReport}. Therefore, existing hardware-based detection and correction techniques will be insufficient to cope with all memory errors. Silent data corruptions (SDC) are particularly problematic because they cause incorrect data to be delivered to applications without an error notification being raised.         

From the perspective of an HPC application, there is an expectation of reliable program execution regardless of how unreliable a system is. However, for various scientific applications, not every computation needs to be performed reliably as long as specific critical computations and their input and result data are guaranteed to be highly reliable. Different variables in a program may exhibit different vulnerabilities to transient faults. Many classes of scientific algorithms are able to leverage such an expectation of selective robustness to continue to make progress towards an acceptable outcome in the presence of some system unreliability \cite{DeBardeleben:2009}. In this paper, we explore the use of selective reliability through a region-based approach to memory allocation. This approach enables the creation of software-enabled robust memory regions called \textit{havens}. We guarantee highly-reliable behavior for these regions through comprehensive error protection based on a lightweight software-based parity protection scheme. Havens permit grouping program objects with similar robustness requirements. Based on this model for memory management, the HPC applications may develop resiliency strategies that trade-off application performance with reliability under program control, by declaring certain program data objects and computations to be highly reliable. Using accelerated fault-injection experiments, we measure the fault coverage of our approach, which confirms that havens provide high levels of fault protection for critical program objects with limited overhead to the overall application performance. 

The rest of the paper is organized as follows: Section \ref{sec:Robust-Regions} motivates the concept and describes its important features. Section \ref{sec:Interfaces} describes abstract haven interfaces, while Section \ref{sec:Implementation} describes our prototype implementation of the interfaces and the parity-based protection algorithm. Section \ref{sec:Experiments} presents the experimental evaluation and results.

\begin{figure*} 
\centering
\includegraphics[width=140mm,height=20mm]{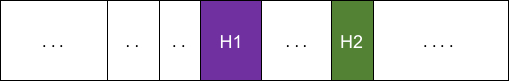}
\caption{Havens-based Memory Management}
\label{Fig:HavensAddressSpace}
\end{figure*}

\section{Resilience-driven Memory Management}
\label{sec:Robust-Regions}

In the region-based memory management scheme for typed, high-level programming languages, the runtime memory is partitioned into \textit{regions} and the assignment of program objects to regions is statically decided by the compiler. The memory allocation and deallocation is statically specified in the program at compile time and the program is annotated to include operations for managing regions \cite{Tofte:1994}. The approach was proposed as an alternative to garbage collection in the context of declarative programming languages. The scheme was improved by relaxing the requirement for region lifetimes to be lexical \cite{Aiken:1995}. However, the efficacy of using static memory management decisions in place of runtime garbage collection has been widely debated.

We propose an adaptation of the region-based memory management concept in which our primary aim is to provide error resilient memory regions for HPC applications. Based on the insight that HPC applications dominantly use imperative programming paradigms, rather than declarative programming, we do not limit the concept of havens to any specific language style. Our proposed approach retains the core concept of well-defined regions for program objects. However, rather than assigning objects to regions solely based on their lexical scopes, we relax several constraints of the original concept and define features that make havens more suitable to HPC applications and programming paradigms and provide reliability. The salient features of our resilience-driven, region-based memory management scheme are:
\begin{itemize}
 \item Havens provide robust containers in which program objects may be allocated. The memory region is protected by a predefined error detection and/or correction scheme that is agnostic to the algorithm features or structure of the data objects placed in havens. 
 \item The havens enable separation between the memory allocation and the implementation of the robustness scheme. In theory, a region can be made robust using various protection mechanisms, e.g., replication, parity, hashing, etc. As illustrated in Figure \ref{Fig:HavensAddressSpace} various regions in the program memory may be offered coverage using separate protection schemes. Since each of these methods carry a different level of overhead, for an application this enables the creation of logical grouping of objects that require similar robustness characteristics. 
 \item Havens enable applications to exert fine-grained control on the resilience properties of individual program objects. Since different havens may have varying guarantees of reliability, based on the strength of the protection mechanism and its performance overhead, object placement in havens may be driven by the trade-off between criticality of the object to program correctness and the associated overhead. 
 \item The objects in a haven are all freed at once by deleting the entire pool of memory. Therefore, havens enable the association of lifetime to the robust memory regions.   
\end{itemize}

The key advantage of havens-based memory management is that it makes it straightforward for HPC applications to reason about the reliability requirements of programs and specific program fragments.

\section{Using Havens in HPC Applications}
\label{sec:Interfaces}

In this section we introduce the abstract interface between the runtime memory manager and the application program. The overall goal of the havens approach is to enable applications to allocate program objects into regions for which there is a well-defined robustness scheme. Each program object that is resident to a region is provided with error protection. In imperative languages such as C/C++, the memory allocation interface consists of two operations: an \textit{alloc} and \textit{free} which create new heap objects. In our scheme, the application program interacts with a haven manager, which implements the haven abstraction and its error protection mechanism and handles the subdivision of the heap memory.

\begin{figure}
\centering
\includegraphics[width=60mm,height=80mm]{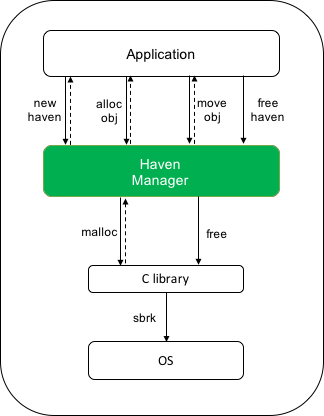}
\caption{Havens Interfaces}
\label{Fig:HavensAbstractInterface}
\end{figure}
\subsection{Creation}
The havens are part of the heap where memory is allocated. When the region is first created, no memory is allocated. The creation is requested by an application using an abstract interface to the haven manager. When the application requests the creation of a new region, the haven manager returns a handle to the memory region, which is to be used in subsequent operations. The abstract interface needs to provide a \texttt{\_\_haven\_create\_\_} primitive for haven creation. Also, the robustness technique may be explicitly specified for the memory region. Upon creation, the haven is an abstract entity for which an application expresses an expectation of reliability. 

\subsection{Allocation}
An application requests allocation of memory within a region using the handle to the haven. This entails allocation of a block consisting of a specified amount of memory. The abstract interface must provide an interface \texttt{\_\_haven\_alloc\_\_}, to which the application specifies a handle to the haven of which the new memory block will be a part of. The haven manager maintains state for the robustness feature of the memory region and the allocation requests initialization of such state.

\subsection{Deallocation}
When the haven is destroyed, all memory blocks allocated in the region are deallocated and the memory is available for reuse. Since this is the only way to deallocate memory in the region model, no further memory allocation requests are valid. Therefore, all the program objects located in a particular region are all simultaneously deallocated. The robustness characteristic of the haven and the state maintained by the haven manager is also destroyed. The abstract interface must provide a \texttt{\_\_haven\_destroy\_\_} primitive, which uses the haven handle. 

\subsection{Read/Write}
The havens interface must provide \texttt{\_\_haven\_read\_\_} and \texttt{\_\_haven\_write\_\_} primitives. The read primitive implicitly detects whether the object is in error state, whenever the object is referenced. The write primitive updates any state maintained by the haven's resilience mechanism, in addition updating the object itself. 

\subsection{Relaxation}
Among the key features of the havens-based memory management is the association of reliability lifetime with the memory contained in the haven. Many HPC applications can naturally divide their work into nearly independent pieces or phases for which the reliability requirement may be more or less stringent. The error protection scheme may be turned on and off based on the needs of the application program. The abstract interface must provide primitives \texttt{\_\_haven\_relax\_\_} and \texttt{\_\_haven\_robust\_\_} in order to relax and impose the protection scheme on the memory region.

\section{Implementation}
\label{sec:Implementation}

The description of the haven interfaces in the previous section is intentionally abstract and language-independent, and a variety of implementations for these interfaces are possible. To simplify porting existing HPC applications to use havens, our initial prototype implementation provides a minimal set of haven library interface calls, which will require limited changes to application source codes.  

\subsection{Basic Haven Operations}
The heap is divided into fixed-size pages, and each new haven creation is aligned on a page-size boundary. The haven manager maintains a linked list of these pages. The library provides concrete functions for each of the primitives that enable basic haven operations. The library provides a \texttt{haven}-type that is used to create handle objects, and which contains the bounds of page addresses that make up the memory region. An identifier is bound to a haven handle and the handle to the haven is passed as argument to the allocation and destruction functions. The functions \texttt{\_\_hmalloc()} and \texttt{\_\_hnew()} implement the abstraction \texttt{\_\_haven\_create()} for the allocation of objects into the associated region. The implementation of the haven system imposes no changes on the representation of pointers and permits the access to individual objects in the havens using pointer operations. However, we only support per-region allocation and deallocation and therefore per-object deallocation is an illegal operation. The \texttt{\_\_hdestroy()} operation concatenates all the haven's page list to a global list of free pages. The present implementation does not support techniques that prevent dangling-pointer dereferences and require the programmer to ensure safety when haven deallocation function is called. 

\subsection{Parity-based Protection}
\begin{figure} [h]
\centering
\includegraphics[width=\linewidth,height=60mm]{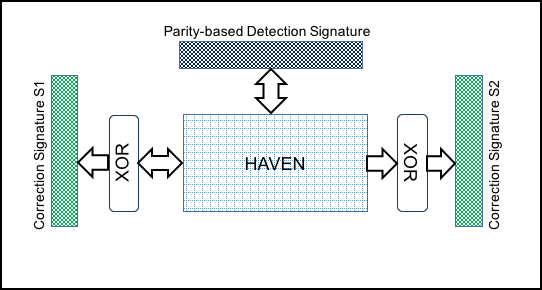}
\caption{Parity-based Protection for Haven Memory}
\label{Fig:ParityProtection}
\end{figure}
The goal of the error protection mechanism is to provide comprehensive fault coverage for the memory of an entire haven. In order to implement a generic scheme that is agnostic to the actual program data structure, yet carries low performance overhead, we employ a parity-based detection/correction mechanism. The region is the unit of memory over which the library applies the parity-based detection/correction.  

The Figure \ref{Fig:ParityProtection} illustrates the implementation of the parity-based detection/correction scheme applied to the haven. The implementation is based on the bit-level representation of the memory in the haven and is independent of the higher-level data type. The haven manager maintains a pair of correction signatures for each memory region, which are of word length, and an additional word length detection signature per 64 words in the memory block in the region. The detection signature contains one parity bit per word in the memory region. As memory is allocated for the region and initialized, the correction signature S1 retains the XOR of all words that are written to the memory region. We apply an XOR operation on every word that is updated in the memory region and the correction signature S2. Silent data corruptions or multi-bit errors are detected based on the state of the detection signature containing a parity bit for each word contained in the memory region. 

When the parity signature for a memory location in the haven detects a parity violation, the location of the corrupted memory word may be identified. The value at the corrupted memory location may be recovered using the XOR signature words. The XOR of the two signatures S1 and S2 equals the XOR of all the uncorrupted locations in the haven. The corrupted value in the memory region is recovered by performing an XOR operation on the remaining words in the haven with the XOR of the two signatures S1 and S2. The recovered value overwrites the corrupted value, and the detection signature is recomputed. Using this correction scheme, multibit corruptions may be recovered from (unlike hardware ECC, which only offers single-bit error correction and double-bit error detection). Each of these of detection/recovery operations are transparent to the application. This parity-based protection is an adaptation of an erasure code and it maintains very limited state for the detection and correction capabilities and therefore carries very little space overhead in comparison to other software-based schemes such as software-based ECC and checksums schemes. The detection is a constant time operation while the recovery is a O(n) operation based on the size of the haven.  

\section{Experimental Evaluation}
\label{sec:Experiments}

\begin{figure*} 
\centering
\includegraphics[width=\linewidth,height=65mm]{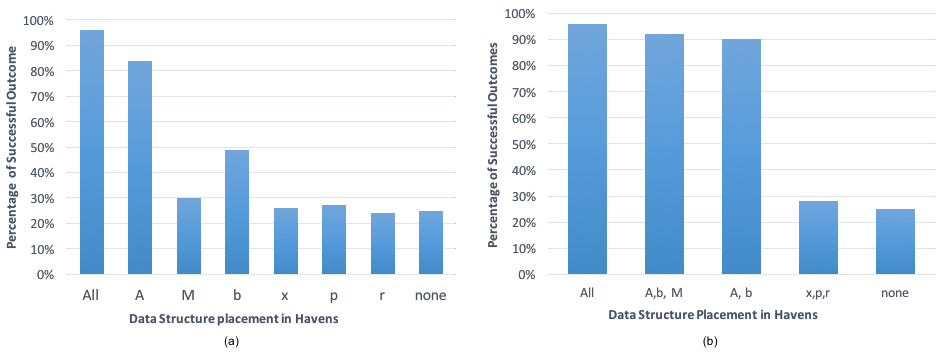}
\caption{Impact of havens' fault coverage on application outcomes}
\label{Fig:ResultsFI}
\end{figure*}

\begin{figure*} 
\centering
\includegraphics[width=\linewidth,height=65mm]{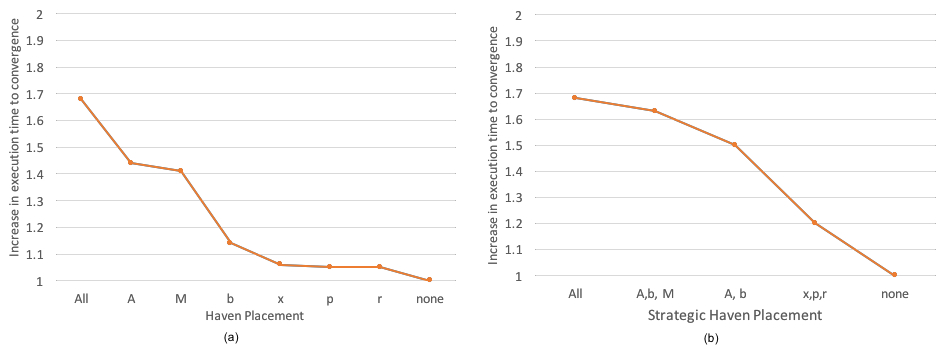}
\caption{Performance overheads of havens}
\label{Fig:ResultsPerf}
\end{figure*}

\subsection{Methodology}
The fault model for these experiments is transient errors, particularly silent data corruptions. Therefore, we have designed an accelerated fault injection system in which we inject multiple random faults into the address space of a program under execution. We perturb multiple bits of the selected injection site and no error notification is raised. The fault location and distribution of fault intervals is randomized, and applications experience at least 5 faults per application run.  

For the evaluation of the haven-based memory management system, we modify a conjugate gradient code to include the havens library API calls. We use a pre-conditioned iterative CG algorithm, and we validate the outcome of the solver with a solution produced using a direct solver. In the CG algorithm, which solves a system of linear equations A.x = b, the application allocates the matrix A, the vector b and the solution vector x. Additionally, the conjugate vectors p and the residual vector r are referenced during each iteration of the algorithm. We perform two sets of experiments, in both of which the various objects in the CG code are allocated using havens. In the first experiment, we evaluate the CG application under accelerated fault injections. For each experiment run, we allocate only one structure using the haven, while the remaining structures are allocated using the standard memory allocation interfaces.   
  
From an error recovery perspective, the application state may be classified into three types of state: the static state, which represents variables initialized at the start and remain unmodified during the computation; the dynamic state, which represents the solution data structures that are continuously updated; and, the computational environment, which represents the program code, environment variable, pointer variables, etc. In the second set of experiments, this classification forms the basis of selecting which data structures in the CG code are allocated into havens. We evaluate the following combinations of CG objects' placements: (i) allocation of only the static state, i.e., the matrix A and vector B, the preconditioner M into havens, while the dynamic state, i.e., all the solution vectors, are allocated using standard memory allocation functions; (ii) allocation of only matrix A and vector B into havens; (iii) only the dynamic state is provided fault coverage using havens.     

\subsection{Results}
Providing fault coverage to all the CG's variables results in a success rate in the excess of 90\% for the accelerated fault injection experiments, as shown in Figure \ref{Fig:ResultsFI}(a). The operand matrix A occupies a dominant part of the solver's memory, occupying over 50\% of the active address space, whereas the solution vector x, the conjugate vectors p and the residual vector r and the preconditioner matrix M account for the remaining space. The preconditioner matrix M demonstrates lower sensitivity to the silent errors, as do the vectors x, p, r. This is because silent errors in the operand matrix A or vector b fundamentally changes the linear system being solved. Even if the CG solver converges to a solution, it may be significantly different from a correct solution. The allocation of these structures into havens provides a substantially higher resilient behavior in terms of completion rates of the CG algorithm. The algorithm is naturally resilient to silent corruptions in the solution vectors, and the results suggest that there is little incentive to allocate these vectors into havens. The preconditioning matrix is also tolerant to errors, but may sometimes require a substantially higher number of iterations to converge to a solution. The performance overhead of using havens to the convergence time of the CG code (Figure \ref{Fig:ResultsPerf}(b)) is higher for higher coverage, but correlating the coverage results demonstrates that higher coverage for low sensitivity program state does not translate to higher application resilience. 

The results of the evaluation of the strategic allocation of program objects using havens (Figure \ref{Fig:ResultsFI}(b)), and their correlation with the performance overhead analysis (Figure \ref{Fig:ResultsPerf}(b)), shows that the allocation of only the matrix A and vector B yields the best return in terms of completion success rate of the application. The remaining combinations of allocation using havens provides diminishing returns. The results indicate that the CG solver suits the model of selective reliability, in which the various program objects show different sensitivities to errors and impact on the outcome of the application. The results indicate that placing only such critical objects in the parity-protected havens yields higher probability of successful outcome and we are able to avoid additional overhead to performance incurred in protecting lesser important program objects.

\section{Related Work}
\label{sec:RelatedWork}

Reliable memory systems have been implemented using hardware-based encoding techniques such as ECC \cite{Bossen:1980}\cite{Chen:1984}. Memory scrubbing techniques were proposed in order to increase memory reliability in noisy environments \cite{Saleh:1990}, which entail periodically reading all memory lines, correcting any bits through ECC, and writing corrected data back to the same location. The use of additional encoding bits for stronger error detection and correction codes adds overheads to the area, energy consumption, and increases the memory access latencies. Other approaches have focused on designing fault tolerant memory cells through cell-hardening techniques \cite{Vargas:1994} \cite{Calin:1996}. The partial protection of cache memory has been accomplished by partitioning sensitive and non-sensitive data \cite{Lee:2006}, and through dynamic assignment of protection priority to error-prone cache lines \cite{Kim:1999}. Partial protection was also explored through a reliability-aware data placement scheme that provided stronger fault protection for only the program code and data sections \cite{Mehrara:2008}.

Software-based selective protection may be accomplished through compiler-directed techniques, in which all statements relevant to the computation of specific variables are extracted using static program analysis, and placed in protection domains \cite{Chen:2005}. Programming constructs called containment domains \cite{Chung:2011} offer transactional semantics for programmer-defined computational scope, where as Rolex \cite{Hukerikar:2016} offers langauge-based extensions that support various resilience semantics on application data and computations. Global View Resilience (GVR) supports reliability of application data by providing an interface for applications to maintain version-based snapshots of the data \cite{Chien:2015}. In support of fault tolerance of dynamic memory allocation, the \texttt{malloc\_failable} may be used by the application to allocate memory on the heap; callback functions are used to handle error recovery of the memory block \cite{Bridges:2011}. Much research has also been done on algorithm-based fault tolerance (ABFT) methods \cite{Huang:1984}\cite{Bosilca:2008}. These approaches apply encoding on the application data structures to detect and recover errors in them. However, such approaches require significant modifications to the application source code to incorporate the fault detection and correction capabilities into the algorithm. The notion of selective reliability was demonstrated as a viable solution for iterative methods for linear solvers, in which the GMRES algorithm is partitioned into reliable and unreliable phases \cite{Hoemmen:2011}.

\section{Conclusion}
\label{sec:Conclusion}

Future HPC systems may not be able to offer complete hardware-based protection mechanisms for memory errors due to constraints of cost and energy. Due to their frequency, which is only expected to increase in future systems, it is imperative to protect critical program objects from errors that are uncorrectable by hardware schemes, and from silent data corruptions. In this paper we presented a software-driven fault tolerance technique for program memory, which extends the concept of region-based memory management. HPC applications may identify the most critical memory objects and place them in robust regions, called havens. The havens are protected using a software-implemented, lightweight parity-based error detection and correction scheme, which provides transparent protection against memory errors without relying on any hardware-based detection/correction schemes. Our solution is agnostic to the algorithm design, and is applicable to a variety of HPC applications that require selective reliability. Our results show that providing such explicitly robust memory havens proves to be an effective way of providing low-cost fault tolerance for applications running on unreliable system fabrics.


\bibliographystyle{IEEEtran}
\bibliography{IEEEabrv,main}

\end{document}